\documentclass[submission,copyright,creativecommons]{eptcs}
\providecommand{\event}{VPT 2018} 
\usepackage{breakurl}             
\usepackage{underscore}           
\usepackage{algorithm2e}          
\usepackage{tabularx}             
\usepackage{lmodern}
\usepackage{graphicx}
\usepackage{amsmath}
\usepackage{amssymb}
\usepackage{tikz}
\usetikzlibrary{calc}

\newcommand{\PR}{\mathit{Rank}}
\newcommand{\seq}[1]{\langle#1\rangle}

\newcommand{\acc}{\mathit{acc}}
\newcommand{\xs}{\mathit{xs}}
\newcommand{\xss}{\mathit{xss}}
\newcommand{\vs}{\mathit{vs}}
\newcommand{\ys}{\mathit{ys}}

\newcommand{\tmap}{\mathsf{map}}
\newcommand{\tzip}[2]{\mathsf{zip} (#1,#2)}
\newcommand{\titer}{\mathsf{iter}}

\newcommand{\tfold}{\mathsf{fold}}

\newcommand{\tunit}{\mathsf{unit}}

\newcommand{\trange}{\mathsf{range}}
\newcommand{\tlength}{\mathsf{length}}

\newcommand{\tgroup}{\mathsf{group}}

\newcommand{\reffig}[1]{Fig.~\ref{fig:#1}}

\newcommand{\introparagraph}[1]{\vspace{-7pt}\paragraph{#1.}}

\newcommand{\IL}{\textsf{IL}\xspace}
\newcommand{\FFL}{\textsf{FFL}\xspace}

\newcommand{\pagerank}{\emph{PageRank}\xspace}

\newcommand{\sideconditions}[1]{\textbf{Side conditions:}\ #1}
\DeclareRobustCommand{\rewriterule}[3]{
  {
  \begin{tabular}[t]{l>{\centering\arraybackslash}p{1cm}l}
    #1 & \(\leadsto\) & #2
  \end{tabular}
  \def\temp{#3}\ifx\temp\empty
  \else
  {
  \\[.3em]
  \noindent
  \sideconditions{#3}
  }
  \fi
  }
}

\newcommand{\mapreduce}{\emph{MapReduce}\xspace}
\newcommand{\coq}{\emph{Coq}}

\newcommand{\numrules}{13}

\newcommand{\forloop}{\texttt{for}}
\newcommand{\foreachloop}{\texttt{for}}
\newcommand{\whileloop}{\texttt{while}}

\usepackage{listings}
\usepackage{lstautogobble}
\usepackage{lstlinebgrd}
\usepackage{xcolor}
\usepackage{tabularx}

\definecolor{codehighlight}{HTML}{FBB829}
\newcommand{\colorinrange}[3]{
  \pgfmathparse{#1<=\value{lstnumber} && \value{lstnumber}<=#2}
  \ifnum\pgfmathresult>0
  \color{#3}
  \fi
}

\lstdefinelanguage{IL}{
    keywords={fn, for, Int, Rat, return, var, while, =>},
    sensitive=true, 
    morecomment=[l]{//}, 
    morecomment=[s]{/*}{*/}, 
    morestring=[b]", 
}

\title{Proving Equivalence Between Imperative and MapReduce
  Implementations Using Program Transformations}
\author{Bernhard Beckert \quad Timo Bingmann \quad Moritz Kiefer \\
  Peter Sanders \quad Mattias~Ulbrich \quad Alexander~Weigl
\institute{Institute of Theoretical Informatics\\
  Karlsruhe Institute of Technology, Germany}
}
\def\titlerunning{Equivalence Between Imperative and MapReduce
  Implementations}
\def\authorrunning{Beckert, Bingmann, Kiefer, Sanders, Ulbrich, Weigl}

\begin{document}

\maketitle

\begin{abstract} 

  Distributed programs are often formulated in popular functional
  frameworks like \mapreduce{}, \emph{Spark} and \emph{Thrill}, but
  writing efficient algorithms for such frameworks is usually a
  non-trivial task.
  As the costs of running faulty algorithms at scale can be severe, it
  is highly desirable to verify their correctness.

  We propose to employ existing imperative reference implementations
  as specifications for \mapreduce{} implementations. 
  To this end, we present a novel verification approach in which
  equivalence between an imperative and a \mapreduce{} implementation
  is established by a series of program transformations.

  In this paper, we present how the equivalence framework can be used
  to prove equivalence between an imperative implementation of the
  \pagerank{} algorithm and its \mapreduce{} variant.
  The eight individual transformation steps are individually presented
  and explained.
\end{abstract}

\section{Introduction}

Today requirements on the efficiency and scale of computations grow
faster than the capabilities of the hardware on which they are to run.
Frameworks such as \mapreduce{}~\cite{mapreduce}, Spark~\cite{spark}
and Thrill~\cite{Thrill} that distribute the computation workload
amongst many nodes in a cluster, address these challenges by providing
a limited set of operations whose execution is automatically and
transparently parallelised and distributed among the available nodes.

In this paper, we use the term ``\mapreduce{}'' as a placeholder for a
wider range of frameworks. While some frameworks such as Hadoop's
MapReduce~\cite{white2012hadoop} strictly adhere to the two functions
``map'' and ``reduce'', the more recent and widely used distribution
frameworks provide many additional primitives -- for performance
reasons and to make programming more comfortable.

Formulating efficient implementations in such frameworks is a
challenge in itself. The original algorithmic structure of a
corresponding imperative algorithm is often lost during that
translation since imperative constructs do not translate directly to
the provided primitives. Significant algorithmic design effort must be
invested to come up with good \mapreduce{} implementations, and flaws
are easily introduced during the process.

The approach we proposed in our previous work~\cite{arxiv} and will refine and apply in
this paper supports algorithm engineers in the correct design of
\mapreduce{} implementations by providing a transformation framework
with which it is possible to interactively and iteratively translate a
given imperative implementation into an efficient one that operates
within a \mapreduce{} framework.

%
The framework is thus a verification framework to prove the
\emph{behavioural equivalence} between an imperative algorithm and its
\mapreduce{} counterpart.
Due to often considerable structural differences between the two
programming paradigms, our approach is interactive: It requires the
specification of intermediate programs to guide the translation
process.
While it is not in the focus of this publication, our approach is
designed to have a high potential for automation: The required
interaction is designed to be as high-level as possible. The rules are
designed such that their side conditions can be proved automatically,
and pattern matching can be used to allow for a more flexible
specification of intermediate steps.


We present an approach based on program transformation rules with
which a \mapreduce{} implementation of an algorithm can be proved
equivalent to an imperative implementation.
From an extensive analysis of the example set of the framework Thrill,
we were able to identify a set of \numrules{}
transformation rules such that a chain of rule applications from this
set is likely to succeed in showing the equivalence between an
imperative and a functional implementation.

We describe a workflow for integrating this approach with existing
interactive theorem provers. We have successfully implemented the
approach as a prototype within the interactive theorem prover
\coq{}~\cite{Coq}.

The main contribution of this paper is the demonstration of the
abilities of the framework to establish equivalence between imperative
and \mapreduce{} implementations.
We do this (1)~by motivating and thoroughly explaining the rationales
and the nature of the transformation rules and (2)~by reporting on the
successful application of the framework to a relevant non-trivial
case study.
We have chosen the \pagerank{} algorithm as the demonstrative example
since it is one of the original and best known \mapreduce application
cases.

\introparagraph{Overview of the approach}

\begin{figure}[t]
  \centering
  \includegraphics{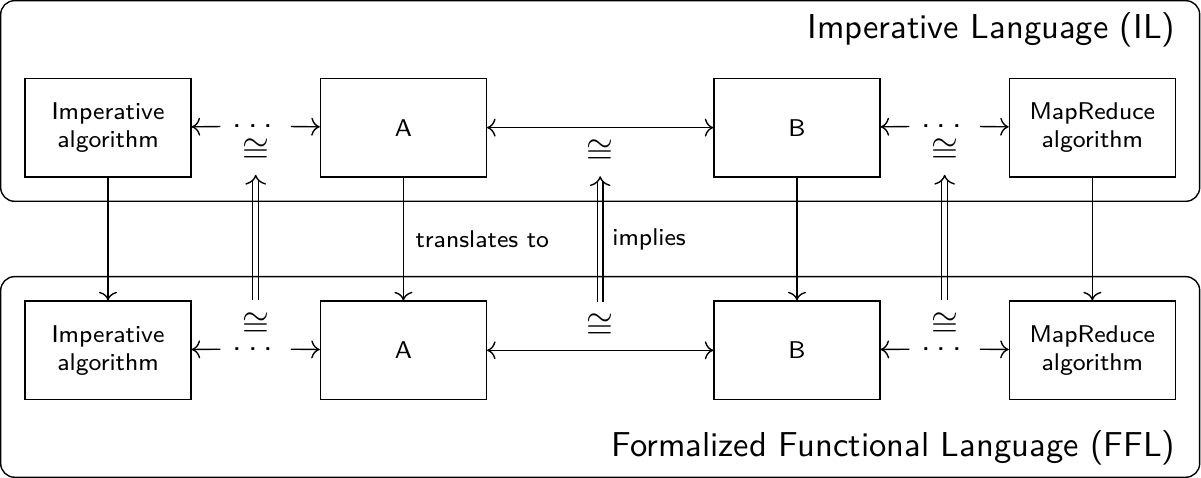}
  \caption{Chain of equivalent programs is translated into formalised functional language}\label{fig:approach-overview}
\end{figure}

The main challenge in proving the equivalence of an imperative and a
\mapreduce{} algorithm lies in the potentially large structural
difference between the two algorithms.
To deal with this, the equivalence of imperative and \mapreduce{}
algorithms is not shown in one step, but as a succession of
equivalence proofs for structurally closer program versions.

To this end, we require that the translation of the algorithm is
broken down (by the user) into a chain of intermediate programs. For
each pair of neighbouring programs in this chain, the difference is
comparatively small such that the pair can be used as start and end
point of a single program transformation step induced by a rule
application.

The approach uses two programming languages: One is the imperative input language
(\IL) in which the imperative algorithm, the intermediate programs, as
well as the target \mapreduce{} implementation are stated. 
Besides imperative language constructs, \IL{} supports
\mapreduce{} primitives. 

Each program specified in the high-level imperative language is then
automatically translated into the formalised functional
language~(\FFL).
The program transformations and equivalence proofs
operate on programs in this functional language.
The translation of the original, the intermediate and the
\mapreduce{} programs form a chain of programs.
For each pair of neighbouring programs in the chain, a proof obligation
is generated that requires proving their equivalence. These proof
obligations are then discharged independently of each other.
%
Since, by construction, the semantics
of \IL{} programs is the same as that of corresponding \FFL{} programs, the
equivalence of two \IL{} programs follows from the equivalence of their
translations to \FFL{}.
An overview of this process can be seen in \reffig{approach-overview}.
Figure~\ref{fig:nonlocal-transformation} shows two example \IL{} programs for
calculating the element-wise sum of two arrays. 

The implementation of our approach based on the \coq{} theorem prover has only
limited proof automation and still requires a significant amount of
interactive proofs.
We are convinced, however, that our approach can be extended such that it
becomes highly automatised and only few user interactions or none at all are
required -- besides providing the intermediate programs. 
%
To enable this automation, one of the design goals of the approach was to
make the matching of the rewrite rules as flexible as possible.
Further challenges
include the extension of our approach to features such as references and
aliasing which are commonly found in imperative languages.



\introparagraph{Structure of the paper}

The remainder of the paper is structured as follows:
After introducing the supported programming languages \IL and \FFL in
Sect.~\ref{sec:foundations}, a description of the two different kinds
of transformation rules applied in the framework follows in
Sect.~\ref{sec:rules}. The case study on the \pagerank{} example is
conducted in Sect.~\ref{sec:example:pagerank}. After a report on
related work in Sect.~\ref{sec:relatedwork}, the paper is wrapped up
with conclusions in Sect.~\ref{sec:conclusion}.

\section{Foundations}
\label{sec:foundations}

This section introduces the programming language~\IL used to
formulate the programs, and briefly describes the language~\FFL{}
used for the proofs.

The high-level imperative programming language~\IL{} is based on a
while language. Besides the usual \texttt{while} loop constructor, it
possesses a \foreachloop{} loop constructor which allows traversal
of the entries of an array.
The supported types are integers (\texttt{Int}), rationals\footnote{In
  the current implementation, rationals are implemented using integers
  with the operators being uninterpreted function symbols.}
(\texttt{Rat}), Booleans (\texttt{Bool}), fixed length arrays
(\texttt{[$T$]}) and sum (\texttt{$T_1$ + $T_2$}) and product types
(\texttt{$T_1$ * $T_2$}). Since arrays are an important basic data type
for \mapreduce, a number of functions are provided to operate on this
data type. 
Table~\ref{tab:il-funcs} lists the \IL-functions relevant for this
paper.
Besides the imperative language constructs, \IL{} supports a number of
\mapreduce{} primitives. In particular, a lambda abstraction for a
variable $v$ of type $T$ over an expression $e$ can be used and is
written as \mbox{\texttt{($v$\,:\,$T$)\,=>\,$e$}.}  \IL{} does not support
recursion.

\begin{table}[t]
  \centering {\footnotesize
  \begin{tabularx}{.9\textwidth}{lX}
    \hline
    \textbf{Function} & \textbf{Explanation}\\
    \hline
    \texttt{replicate($n$, $x$)} &
    returns an array of length $n$ whose entries hold value~$x$.\\
    \texttt{range($a$, $b$)} &
    returns an array containing the values $\seq{a, \dots, b-1}$.\\
    \texttt{zip($\xs$, $\ys$)} &
    returns for two arrays of equal length an array of pairs
    containing a value of each array.\\
    \texttt{map($f$, $\xs$)} &
    returns an array of the same length as $\xs$ that contains
    the result of applying function $f$ to the values in $\xs$.\\
    \texttt{fst($p$)}, \texttt{snd($p$)}&
    returns the first (second) component of a pair $p$ of values.\\
    \texttt{group($\xs$)}& transforms a list of key-value pairs into a
    list of pairs of a key and a list of all values associated with that key.\\
    \texttt{concat($\xss$)}&
    returns the concatenation of all arrays in the array of arrays $\xss$.\\
    \texttt{flatMap($f$, $\xss$)} & = \texttt{map($f$, concat($\xss$))}\\
    \texttt{reduceByKey($f$, $i$, $\xs$)} & = \texttt{map(($k$,$\vs$) => ($k$, fold($f$,$i$,$\vs$)), group($\xs$)) } \\
    \hline
  \end{tabularx}}
  \caption{Relevant built-in functions of \IL.}
  \label{tab:il-funcs}
\end{table}

Given that allegedly \mapreduce{} programs tend to be more of a
functional than an imperative nature, it might seem odd that we use
\IL{} also for specifying the \mapreduce{} algorithm and not a
functional language.
However, most existing \mapreduce{} frameworks are not implemented as
separate languages, but as frameworks built on top of imperative
languages. This implies that the sequential imperative constructs of
the host language can also be found in \mapreduce programs. Sequential
parts of \mapreduce{} algorithms are realised using imperative
programming features, while the computational, distributed parts are
composed using the \mapreduce primitives.
Figure~\ref{fig:nonlocal-transformation} shows two behaviourally
equivalent implementations of a routine that computes the sum of the
entries of two \texttt{Int}-arrays.

\begin{figure}[t]
  \centering
  \begin{minipage}[t]{.48\textwidth}
\begin{lstlisting}
fn SumArrays(xs: [Int], ys: [Int]) {
  var sum := replicate(n, 0);
  for(i : range(0, length(xs))) {
    sum[i] := xs[i] + ys[i];
  }
  return sum;
}
\end{lstlisting}
  \end{minipage}
  \hfill
  \begin{minipage}[t]{.48\textwidth}
\begin{lstlisting}
fn SumArraysZipped(xs: [Int], ys: [Int]) {
  var sum := replicate(n, 0);
  zipped := zip(xs, ys);
  for(i : range(0, length(xs))) {
    sum[i] := fst(zipped[i]) + snd(zipped[i]);
  }
  return sum;
}
\end{lstlisting}
  \end{minipage}
  \caption{Two \IL{} programs which calculate the element-wise sum of
    two arrays.}\label{fig:nonlocal-transformation}
\end{figure}

The programs specified in \IL{} are then automatically translated into
\FFL, the functional language based on $\lambda$-calculus in which the
equivalence proofs by program transformation are conducted.
We follow the work by Radoi~et~al.~\cite{translatingimperative} and
use a simply typed lambda calculus extended by the theories of sums,
products, and arrays.
Moreover, to allow the translation of both imperative and \mapreduce \IL code
into \FFL, the language also contains constructs for loop iteration
and the programming primitives usually found in \mapreduce{}
frameworks.
\FFL is formally defined as a theory in the theorem prover \coq{} which
allowed us to conduct correctness proofs on the rewrite rules.

Without going into the intricacies of the details of the translation
between the formalisms, the idea is as follows: Any \IL statement
becomes a (higher order) term in \FFL in which the currently available
local variables $\acc$ make up the $\lambda$-abstracted variables.
The two primitives $\titer$ and $\tfold$ serve as direct
translation targets for imperative loops. 
The $\tfold$ function is used to translate bounded \foreachloop{}
loops into \FFL. The iterator loop \texttt{for(x\,:\,xs) \char`\{{} f
  \char`\}} in \IL is translated into \FFL as the expression
$\lambda \acc.\ \tfold(\hat f, \acc, \hat{\xs})$ in which $\hat f$ and
$\hat xs$  are the \FFL-translations of \texttt{f} and \texttt{xs}.
This starts with the initial loop state $\acc$ and iterates over each
value of the array $\hat\xs$ updating the loop state by applying
$\hat f$.
The more general \texttt{while} loops cannot be translated using \texttt{fold} since that
always has bounded number of iterations.
Instead, \texttt{while} is translated using the $\titer$ fixed point operator.
The loop \texttt{while(c) \char`\{{} f \char`\}} translates as
$\titer(\lambda \acc.\ \mathsf{if}\ c(\acc)\ \mathsf{then\ inr} \hat
f(\acc)\ \mathsf{else\ inl}\ \tunit)$ and
is evaluated by repeatedly applying $\hat f$ to the loop state until
$\hat f$ returns $\tunit$ to indicate termination.
$\titer$ is a partial function; if the loop does not terminate it does
not yield a value.
The semantics of \FFL is defined as a bigstep semantics. A program
together with its input parameters are reduced to a value.  Ill-typed
or non-terminating programs do not reduce to a value.
Details on the design of \IL and \FFL can be found elsewhere~\cite{arxiv}.

The design of \FFL's core follows the ideas of
Chen~et~al.~\cite{sparkspecification} who describe how to reduce the
large number of primitives provided by \mapreduce{} frameworks.

\section{Transformation Rules}
\label{sec:rules}

For the examples shipped with the \mapreduce{} framework
Thrill, we analysed how the steps of a transformation of an algorithm into
\mapreduce{} would look and detected typical recurring patterns for
steps in a transformation.
We were able to identify two different categories of transformation
rules that are needed for the process:
\begin{enumerate}
\item Local, \emph{context-independent rewrite rules} with which a
  statement of the program can be replaced with a semantically
  equivalent statement. Such rules may have side conditions on the
  parts on which they apply, but they cannot restrict the application context 
  (the statements enclosing the part to be exchanged).
\item \emph{Context-dependent equivalence rules} that cannot replace a
  statement without considering the context. Some transformations are
  only valid equivalence replacements within their surrounding
  context. These are not pattern-driven rewrite rules, but follow a
  deductive reasoning pattern that proves equivalence of a piece of
  code locally.
\end{enumerate}

Context-independent rules are good for changing the control flow of
the program. In Sect.~\ref{sec:trans1}, we will encounter an example of
a rule which replaces a loop by an equivalent \texttt{map}
expression. The data flow, while differently processed, remains the
same.
The context-independent rules are a powerful tool to bridge larger
differences in the control structure between the two programming
paradigms. These changes must be anticipated beforehand and cannot be
detected and proved on the spot. We identified a total of \numrules{} rules that
allow us to transform imperative constructs into \mapreduce{} primitives. 
(See App.~B of~\cite{arxiv} for a complete list.)
Their
rigid search patterns
make context-independent rules less flexible in their application.

Context-dependent rules, on the other hand, are suited for
transforming a program into a structurally similar program (they do
not/little change the control flow); the data flow may be altered
however using such rules. It is comprehensible that a change in the
data representation is an aspect which cannot be considered locally,
but requires analysing the whole program.

The collection of rewrite rules for context-independent replacements
comprises various different patterns.
The context-dependent case is different. There exists one single rule
which can be instantiated for arbitrary coupling predicates and is
thereby highly adaptable. We employ relational reasoning using product
programs~\cite{productprograms} to show connecting properties. The
rule bases on the observation that loops in the two compared programs
need not be considered separately but can be treated simultaneously.
If $x_1$ ($x_2$) are the variables of the first (second) program, and
if the conditions $c_1$ and $c_2$, as well as the loop bodies $b_1$ and
$b_2$ refer only to variables of the respective program, then the
validity of the Hoare triple
\begin{equation}
  \label{eq:hoare-1}
  \{ x_1 = x_2 \} \texttt{ while($c_1$) \char`\{{} $b_1$ \char`\}{} ; 
    while($c_2$) \char`\{{} $b_2$ \char`\}{} }  \{ x_1 = x_2 \}
\end{equation}
is implied by the validity of
\begin{equation}\label{eq:hoare-2}
  \{ x_1 = x_2 \}  \texttt{ while($c_1$) \char`\{{} $b_1$; $b_2$; assert $c_1$=$c_2$; \char`\}{} }  \{ x_1 = x_2 \}\rlap{ \enspace.}
\end{equation}

Condition~\eqref{eq:hoare-1} expresses that given equal inputs, the
two loop programs terminate in the same final
state. Condition~\eqref{eq:hoare-2} manages to
express\footnote{\eqref{eq:hoare-2} is stronger than
  \eqref{eq:hoare-1} in general, but is equivalent in case both loops
  are guaranteed to have the same number of iterations.} this with a
single loop.
This gives us the chance to prove equivalence using a single \emph{coupling
invariant} that relates to both program states.
To show equivalence for context-dependent cases, the specification of
a relational coupling invariant with which the consistency of the data
can be shown is required.

An example of two programs which are equivalent with similar control
structure, yet different data representation, has already been
presented in Fig.~\ref{fig:nonlocal-transformation}. Both programs can
be shown equivalent by means of the coupling invariant
\mbox{\(\mathtt{sum}_{1} = \mathtt{sum}_{2} \wedge \mathtt{zipped_{2}}
  = \tzip{\xs_{1}}{\ys_{1}}\)} where the subscripts indicate which program a variable belongs to.
Sect.~\ref{sec:trans2} demonstrates the application of the rule within
the \pagerank{} example.


In the formal approach (also outlined in \cite{arxiv}) and the \coq{}
implementation, all rules are formulated on and operate on the level
of \FFL (which has been designed for exactly this purpose). 
For the sake of better readability we show the corresponding rules
here on the level of \IL and notate the context independent rewrite
rules as
\begin{displaymath}
  \mathtt{program_1} \quad \leadsto \quad \mathtt{program_2} \qquad (\textsl{rulename})
\end{displaymath}
meaning that under the specified side conditions any statement in a
program matching \texttt{program$_1$} can be replaced by the
corresponding instantiation of \texttt{program$_2$}, yielding a
program behaviourally equivalent to original program.

The application of rules are transformations only in a broader
sense. The approach is targeted as interacting with a
user who provides the machine with the intermediate steps and the
rule scheme to be applied. It would likewise be possible to instead
allow specifying which rules must be applied and have the user specify
the instantiations instead of the resulting intermediate programs.

In particular the context-dependent rule is hardly a
rewrite rule due to the deductive nature of the equivalence
proof. It is not astonishing though that the transformation into
\mapreduce{} does not work completely by means of a set of rewrite
patterns since transforming imperative programs to \mapreduce{}
programs is more than simple pattern matching process but requires
some amount of ingenious input to come up with an efficient
implementation.

\newcommand{\prlisting}[1]{Listing~\ref{lst:pr-#1} on page~\pageref{lst:pr-#1}}
\section{Example: PageRank}
\label{sec:example:pagerank}

In this section, we demonstrate our approach by applying it to the
\pagerank{} algorithm. We present all intermediate programs in \IL{}
and explain the transformations and techniques used in the individual
steps. While the implementation executes the actual equivalence proofs on \FFL{}
terms, we only present the \IL{} programs here since these are the
intermediate steps specified by the user.  Where the translation of a
transformation to \FFL{} is not straightforward, we also give an
explanation of the transformation expressed on \FFL{} terms.

\subsection{The algorithm}

\pagerank is the algorithm that Google originally successfully
employed to compute the rank of web pages in their search
engine. This renowned algorithm is actually a special case of sparse matrix-vector multiplication, which has much broader applications in scientific computing.
\pagerank{} is particularly well suited as an example
for a map reduce implementation and is included in the examples
of Thrill and Spark.
While more highly optimized \pagerank algorithms and implementations exist, we present here a simplified version.

The idea behind \pagerank{} is the propagation of reputation of a web page
amongst the pages to which it links. If a page with a high reputation
links to another page, the latter page's reputation thus increases.

The algorithm operates as follows:
\pagerank{} operates on a graph in which the nodes are the pages of the
considered network, and (directed) edges represent links between
pages. Pages are represented as integers, and the 2-dimensional array
\texttt{links} holds the edges of the graph in the form of an adjacency
list: the $i$-th component of \texttt{links} is an array containing
all indices of the pages to which page $i$ links. The result is an
array \texttt{ranks} of rationals that holds the pagerank value of
every page.
The initial rank is set to $\PR_0(p) = \frac{1}{|\mathtt{links}|}$ for
all pages $p$.

In the course of the $k$-th iteration ($k>0$) of the algorithm, the
rank of each link target is updated depending on the pages that link
\emph{to} the page, i.e., the incoming edges in the graph:
\begin{align}
  \Delta_k(p) &= \sum_{(o,p) \in \mathtt{links}} \mathit{\PR}_{k-1}(o) &
  \mathit{\PR}_k(p) &= \delta * \Delta_k(p) + \frac{1-\delta}{|\mathtt{links}|}
  \label{eq:pr}
\end{align}
The factor $\delta \in (0,1)$ is a dampening factor to
limit the effects of an iteration by weighting the influence of the
result of the iteration $\Delta_k(p)$ against the original value
$\PR_0(p) = \frac{1}{|\mathtt{links}|}$.
Our implementation iterates this scheme for a fixed number of times
(\texttt{iterations}).

\prlisting{1} shows a
straightforward imperative \IL implementation of this algorithm that
realises the iterative laws of \eqref{eq:pr} directly. It marks the
starting point of the translation from imperative to \mapreduce
algorithm.
To allow a better comparison between the programs, the programs are
listed next to each other at the end of this section.


\subsection{A context-independent rule application}
\label{sec:trans1}

The first step in the chain of transformations from imperative to
distributed replaces the \foreachloop{} loop used to calculate the
weighted new ranks with a call to \texttt{map}. This is possible since
the values can be computed independently. The \texttt{map}
expression allows computing the dampened values in parallel and can
thereby significantly improve performance. The rewrite rule used here
can be applied to all \foreachloop{} loops that iterate over the index
range of an array where each iteration reads a value from one array at
the respective index, applies a function to it and then writes the
result back to another array at the same index.

\begin{minipage}{.3\linewidth}
\begin{lstlisting}
    for (i : range(0, length(xs))) {
      ys[i] := f(xs[i]);
    }
\end{lstlisting}  
\end{minipage}
\hfill
$\leadsto$
\hfill
\begin{minipage}{.3\linewidth}
\begin{lstlisting}
    ys := map(f, xs);
\end{lstlisting}  
\end{minipage}
(\textsl{map-introduce})

Sufficient conditions for the validity of this context-independent
transformation are that \texttt{f} does not access the index \texttt{i}
directly and that \texttt{xs} and \texttt{ys} have the same length.
The first condition can be checked syntactically while matching the
rule while the second requires a (simple) proof in the context of the
rule application. In our implementation, these proofs are executed in \coq{}.

As mentioned before, \foreachloop{} loops in \IL{} correspond to
\texttt{fold} operations in \FFL{}. The rewrite rule expressed on
\FFL{} thus transforms
$\tfold(\lambda \acc\ i.\ \acc[i := f(\mathit{xs}[i])], \mathit{ys},
\trange(0, \tlength(xs)))$ into $\tmap(f, \mathit{xs})$.

The result of the transformation is shown in \prlisting{2}.  For
convenience, in this and the following listings, the modified part of
the program is highlighted in colour.

\subsection{A context-dependent rule application}
\label{sec:trans2}

In this step, the body of the main \whileloop{} loop is changed to
first combine the \texttt{links} and \texttt{ranks} arrays to an array
\texttt{outRanks} of tuples using the \texttt{zip} primitive. In the
remaining loop body, all references to \texttt{links} and
\texttt{ranks} point to this new array and retrieve the original
values using the pair projection functions \texttt{fst} and
\texttt{snd}. 
The process of rewriting all references does not fit easily into the
rigid structure of the rewrite rules employed in our approach. We thus
resort to using a context-dependent rule using a coupling predicates
to prove equivalence of the last and the new loop body.
Using the coupling predicate 
\begin{displaymath}
  \mathtt{newRanks}_{1} = \mathtt{newRanks}_{2} ~~\wedge~~
  \mathtt{outRanks}_{2} = \mathtt{zip}(\mathtt{links}_{1}, \mathtt{ranks}_{1})
\end{displaymath}
that relates the values in the states of the two programs
(we use the subscript indices \(1\) and \(2\) to refer to variables in the original and the transformed program) we obtain
that the loop bodies have equivalent effects, and hence, that the
programs are equal.

The result of the transformation is shown in \prlisting{3}.

\subsection{Rule \textsl{range-remove}}

In the next transformation the \forloop{} loop which iterates over all
pages as the index range of the array \texttt{links} is replaced by a
\foreachloop{} loop that iterates directly over the elements in
\texttt{outRanks}. The rewrite rule \textsl{range-remove} applied here
can be applied to all \foreachloop{} loops that iterate over the index
range of an array and only use the index to access these array
elements. Again this is a side condition for the rule which can be
checked syntactically during rule matching.

\begin{minipage}{.3\linewidth}
\begin{lstlisting}
    acc := acc0;
    for (i : range(0, length xs)) {
      acc := f(acc, xs[i]);
    }
\end{lstlisting}
\end{minipage}
\hfill
$\leadsto$
\hfill
\begin{minipage}{.3\linewidth}
\begin{lstlisting}
    acc := acc0;
    for (x : xs) {
      acc := f(acc, x);
    }
\end{lstlisting}
\end{minipage}
(\textsl{range-remove})

\noindent The result of the application of rewrite rule is shown in
\prlisting{4}.

Expressed on the level of \FFL{}, this rewrite rule transforms terms of the form \\
\begin{math}
  \tfold(\lambda \acc\ i.\ f(\acc,\xs[i]), \acc_0, \trange(0,\tlength(\xs))
\end{math}
  into
  \begin{math}
    \tfold(\lambda \acc\ x .\ f(\acc,x), \acc_0, \xs)
  \end{math}.

\subsection{Aggregating link information}

The next step is a typical step that can be observed when migrating
algorithms into the \mapreduce{} programming model. A computation is
split into two consecutive steps: one step processing data locally on
individual data points and one step aggregating the results. It can
be anticipated already now that these two steps will become the
\emph{map} and the \emph{reduce} part of the algorithm.

The newly introduced variable \texttt{linksAndContrib} stores the
(locally for each node) computed rank contribution as a list of
tuples. Assume $(\seq{s_1, \ldots, s_n}, r)$ is the $i$-th entry in
the array \texttt{outRanks}. This means that page $i$ links to page $s_j$
for $j < n$ and has a current rank of $r$. After the newly
introduced local computation, the entry becomes the list of pairs
$\seq{(s_1, \frac{r}{|\mathtt{links}|}), \ldots, (s_n,
  \frac{r}{|\mathtt{links}|})}$,
i.e., the rank is distributed to all successor pages and the data is
rearranged with the focus now on the receiving pages.

As in the transformation in Sect.~\ref{sec:trans2}, a
context-dependent transformation is employed to prove equivalence
using the following relational coupling loop invariant:
\begin{displaymath}
  \begin{aligned}
    & \mathtt{newRanks}_{1} = \mathtt{newRanks}_{2} \ \wedge \\
    & \begin{aligned}
      \forall i j.\ & \mathtt{fst}\ \mathtt{linksAndContrib}_{2}[i][j] =
                      (\mathtt{fst}\ \mathtt{outRanks}_{1}[i])[j] \ \wedge \\
           & \mathtt{snd}\ \mathtt{linksAndContrib}_{2}[i][j] = \mathtt{snd}\ \mathtt{outRanks}_{1}[i] / \mathtt{length} (\mathtt{fst}\ \mathtt{outRanks}_{1}[i])
      \end{aligned}
  \end{aligned}
\end{displaymath}

The result of the transformation is shown in \prlisting{5}. 
Note that the nested loops in the highlighted
block no longer perform the computation of the rank updates
(\texttt{snd links_rank / length(fst links_rank)}), but only aggregate
the contribution updates into new ranks.
This transformation is a preparation for collapsing the nested
loops in the next step.

\subsection{Collapsing nested loops}
\label{sec:trans4}

Since the computation of \texttt{contribution} has been moved outside
in the previous step, the iteration variable \texttt{link_contribs} is
now only used as the iterated array in the inner \foreachloop{}
loop. This allows collapsing the nested loops into a single loop using
\texttt{concat}. This rule can always be applied if the iterated value
in the inner \foreachloop{} is the only reference to the values the
outer \foreachloop{} iterates over.

\begin{minipage}{.3\linewidth}
\begin{lstlisting}
    acc := acc0;
    for (xs : xss) {
      for (x : xs) {
        acc := f(acc, x);
      }
    }
\end{lstlisting}
\end{minipage}
\hfill
$\leadsto$
\hfill
\begin{minipage}{.3\linewidth}
\begin{lstlisting}[mathescape]
    acc := acc0;
    for (x : concat(xss)) {
      acc := f(acc, x);
    }

    $\ $
\end{lstlisting}
\end{minipage}
(\textsl{concat-intro})

The program with the two loops collapsed is shown in
\prlisting{6}.

This transformation is succeeded by a step that combines the call to
\texttt{concat} in the \foreachloop{} loop and the \texttt{map}
operation before the loop into a single call to \texttt{flatMap}.
Its result is shown in \prlisting{7}.
In \FFL{}, \texttt{flatMap} is not a builtin primitive but a synonym
for successive calls to \texttt{concat} and \texttt{map}.  This step
is thus one which has visible effects on the level of \IL, but no
impact on the level of \FFL.

\subsection{Towards \mapreduce}

Now we are getting closer to a program that adheres to the
\mapreduce{} programming model. 
The penultimate transformation step restructures the processed data by
grouping all rank updates that affect the same page.
It operates on the array \texttt{newRanks} using the function
\texttt{group}. The updated result is calculated using a combination
of \texttt{map} and \texttt{fold}. The results are then written back
to \texttt{newRanks}. The effects of the rule on the program structure
are more severe than for the other applied transformation rules, yet
this grouping pattern is one that is typically observed in the
\mapreduce{} transformation process and is implemented as a single 
rule for that reason.

The corresponding rewrite rule can be applied to all \foreachloop{}
loops that iterate over an array containing index-value tuples and
update an accumulator based on the old value stored for that index and
the current value:\footnote{The actually implemented version of the
  rule allows \texttt{f} to access not only the values \texttt{vs},
  but also the index \texttt{i} it operates on. See \cite{arxiv} for
  details.}

\noindent
\begin{minipage}{.25\linewidth}
\begin{lstlisting}[mathescape]
acc := acc0;
for ((i,v) : xs) {
  acc[i] := f(acc[i], v);
}

$\ $
\end{lstlisting}
\end{minipage}
\hfill
$\leadsto$
\hfill
\begin{minipage}{.45\linewidth}
\begin{lstlisting}[mathescape]
 acc := acc0;
 var upd := map((i,vs) => fold(f, acc[i], vs), 
                group(acc));
 for (x : concat(xss)) {
   acc := f(acc, x);
 }
\end{lstlisting}
\end{minipage}\hfill
(\textsl{group-intro})


Note that since \texttt{acc0} could store values for indices for which
there are no corresponding tuples in \texttt{xs}, it is necessary to
write the results back to that array instead of simply using the
result from the \texttt{group} operation which would be missing those
entries.

The resulting program is shown in \prlisting{8}.

\subsection{The final \mapreduce implementation}

In the last step, the expression that groups contributions by index
and then sums them up is replaced by the \IL-function
\texttt{reduceByKey} which is also provided by many \mapreduce{}
frameworks.
In the lower level language \FFL{}, however, \texttt{reduceByKey} is
not a primitive function, but a composed expression using $\tmap$,
$\tfold$ and $\tgroup$, such that this step changes the \IL program,
but has no impact on the $\FFL$ level.
The resulting implementation using map reduce constructs is shown in
\prlisting{9}. It is very close to the \mapreduce implementation of
\pagerank{} that is delivered in the example collection of the Thrill
framework.


\lstset{
  basicstyle=\scriptsize\ttfamily\fontsize{8}{8}
\fontseries{l}\selectfont, 
}
\begin{lstlisting}[caption={Original imperative \IL implementation of \pagerank}, label={lst:pr-1}]
fn pageRank(links : [[Int]], dampening : Rat, iterations : Int) -> [Rat] {
  var iter : Int := 0;
  var ranks : [Rat] := replicate(length(links), 1. / length(links));
  while (iter < iterations) {
    var newRanks : [Rat] := replicate(length(links), 0);
    for (pageId : range(0, length(links))) {
      var contribution : Rat := ranks[pageId] / length(links[pageId]);
      for (outgoingId : links[pageId]) {
        newRanks[outgoingId] := newRanks[outgoingId] + contribution;
      }
    }
    for (pageId : range(0, length(links))) {
      ranks[pageId] :=
        dampening * newRanks[pageId] + (1 - dampening) / length(links);
    }
    iter := iter + 1;
  }
  return ranks;
}
\end{lstlisting}

\noindent
\rule{\linewidth}{.5pt}

\begin{lstlisting}[caption={\pagerank{} -- After applying rule \textsl{map-introduce}},linebackgroundcolor=\colorinrange{12}{14}{codehighlight}, label={lst:pr-2}]
fn pageRank(links : [[Int]], dampening : Rat, iterations : Int) -> [Rat] {
  var iter : Int := 0;
  var ranks : [Rat] := replicate(length(links), 1. / length(links));
  while (iter < iterations) {
    var newRanks : [Rat] := replicate(length(links), 0);
    for (pageId : range(0, length(links))) {
      var contribution : Rat := ranks[pageId] / length(links[pageId]);
      for (outgoingId : links[pageId]) {
        newRanks[outgoingId] := newRanks[outgoingId] + contribution;
      }
    }
    ranks :=
      map((rank : Rat) => dampening * rank + (1 - dampening) / length(links),
          newRanks);
    iter := iter + 1;
  }
  return ranks;
}
\end{lstlisting}

\noindent
\rule{\linewidth}{.5pt}

\begin{lstlisting}[caption={\pagerank{} -- After applying a context-dependent rule},linebackgroundcolor=\colorinrange{6}{12}{codehighlight}, label=lst:pr-3]
fn pageRank(links : [[Int]], dampening : Rat, iterations : Int) -> [Rat] {
  var iter : Int := 0;
  var ranks : [Rat] := replicate(length(links), 1 / length(links));
  while (iter < iterations) {
    var newRanks : [Rat] := replicate(length(links), 0);
    var outRanks : [[Int] * Rat] := zip(links, ranks);
    for (pageId : range(0, length(links))) {
      var contribution : Rat := snd outRanks[pageId] / length(fst outRanks[pageId]);
      for (outgoingId : fst outRanks[pageId]) {
        newRanks[outgoingId] := newRanks[outgoingId] + contribution;
      }
    }
    ranks :=
      map((rank : Rat) => dampening * rank + (1 - dampening) / length(links),
          newRanks);
    iter := iter + 1;
  }
  return ranks;
}
\end{lstlisting}

\noindent
\rule{\linewidth}{.5pt}

\begin{lstlisting}[caption={\pagerank{} -- After applying \textsl{range-remove}},linebackgroundcolor=\colorinrange{7}{12}{codehighlight}, label={lst:pr-4}]
fn pageRank(links : [[Int]], dampening : Rat, iterations : Int) -> [Rat] {
  var iter : Int := 0;
  var ranks : [Rat] := replicate(length(links), 1 / length(links));
  while (iter < iterations) {
    var newRanks : [Rat] := replicate(length(links), 0);
    var outRanks : [[Int] * Rat] := zip(links, ranks);
    for (links_rank : outRanks) {
      var contribution : Rat := snd links_rank / length(fst links_rank);
      for (outgoingId : fst links_rank) {
        newRanks[outgoingId] := newRanks[outgoingId] + contribution;
      }
    }
    ranks :=
      map((rank : Rat) => dampening * rank + (1 - dampening) / length(links),
          newRanks);
    iter := iter + 1;
  }
  return ranks;
}
\end{lstlisting}

\noindent
\rule{\linewidth}{.5pt}

\begin{lstlisting}[caption={\pagerank{} -- After aggregating the link information},linebackgroundcolor=\colorinrange{7}{18}{codehighlight}, label={lst:pr-5}]
fn pageRank(links : [[Int]], dampening : Rat, iterations : Int) -> [Rat] {
  var iter : Int := 0;
  var ranks : [Rat] := replicate(length(links), 1 / length(links));
  while (iter < iterations) {
    var newRanks : [Rat] := replicate(length(links), 0);
    var outRanks : [[Int] * Rat] := zip(links, ranks);
    var linksAndContrib : [[Int * Rat]] :=
      map((links_rank : [Int] * Rat) =>
             map((link : Int) =>
                   (link, snd links_rank / length(fst links_rank)),
                 fst links_rank),
          outRanks);
    for (link_contribs : linksAndContrib) {
      for (link_contrib : link_contribs) {
        newRanks[fst link_contrib] :=
          newRanks[fst link_contrib] + snd link_contrib;
      }
    }
    ranks :=
      map((rank : Rat) => dampening * rank + (1 - dampening) / length(links),
          newRanks);
    iter := iter + 1;
  }
  return ranks;
}
\end{lstlisting}

\noindent
\rule{\linewidth}{.5pt}

\begin{lstlisting}[caption={\pagerank{} -- After collapsing nested loops},linebackgroundcolor=\colorinrange{13}{16}{codehighlight}, label={lst:pr-6}]
fn pageRank(links : [[Int]], dampening : Rat, iterations : Int) -> [Rat] {
  var iter : Int := 0;
  var ranks : [Rat] := replicate(length(links), 1 / length(links));
  while (iter < iterations) {
    var newRanks : [Rat] := replicate(length(links), 0);
    var outRanks : [[Int] * Rat] := zip(links, ranks);
    var linksAndContrib : [[Int * Rat]] :=
      map((links_rank : [Int] * Rat) =>
             map((link : Int) =>
                   (link, snd links_rank / length(fst links_rank)),
                 fst links_rank),
          outRanks);
    for (link_contrib : concat(linksAndContrib)) {
      newRanks[fst link_contrib] :=
        newRanks[fst link_contrib] + snd link_contrib;
    }
    ranks :=
      map((rank : Rat) => dampening * rank + (1 - dampening) / length(links),
          newRanks);
    iter := iter + 1;
  }
  return ranks;
}
\end{lstlisting}

\noindent
\rule{\linewidth}{.5pt}

\begin{lstlisting}[caption={\pagerank{} -- After introducing \texttt{flatMap}},linebackgroundcolor=\colorinrange{8}{8}{codehighlight}\colorinrange{13}{13}{codehighlight}, label={lst:pr-7}]
fn pageRank(links : [[Int]], dampening : Rat, iterations : Int) -> [Rat] {
  var iter : Int := 0;
  var ranks : [Rat] := replicate(length(links), 1 / length(links));
  while (iter < iterations) {
    var newRanks : [Rat] := replicate(length(links), 0);
    var outRanks : [[Int] * Rat] := zip(links, ranks);
    var linksAndContrib : [Int * Rat] :=
      flatMap((links_rank : [Int] * Rat) =>
             map((link : Int) =>
                   (link, snd links_rank / length(fst links_rank)),
                 fst links_rank),
          outRanks);
    for (link_contrib : linksAndContrib) {
      newRanks[fst link_contrib] :=
        newRanks[fst link_contrib] + snd link_contrib;
    }
    ranks :=
      map((rank : Rat) => dampening * rank + (1 - dampening) / length(links),
          newRanks);
    iter := iter + 1;
  }
  return ranks;
}
\end{lstlisting}

\noindent
\rule{\linewidth}{.5pt}

\begin{lstlisting}[caption={\pagerank{} -- After grouping the input for receiving pages},linebackgroundcolor=\colorinrange{12}{19}{codehighlight}, label={lst:pr-8}]
fn pageRank(links : [[Int]], dampening : Rat, iterations : Int) -> [Rat] {
  var iter : Int := 0;
  var ranks : [Rat] := replicate(length(links), 1 / length(links));
  while (iter < iterations) {
    var outRanks : [[Int] * Rat] := zip(links, ranks);
    var contribs : [Int * Rat] :=
      flatMap((links_rank : [Int] * Rat) =>
                 map((link : Int) => (link,
                                      snd links_rank / length(fst links_rank)),
                     fst links_rank),
              outRanks);
    var rankUpdates : [Int * Rat] :=
      map((link : Int) (contribs : [Rat]) =>
            (link, fold((x: Rat) (y : Rat) => x + y, 0, contribs)),
          group(contribs));
    var newRanks : [Rat] := replicate(length(links), 0);
    for (link_rank : rankUpdates) {
      newRanks[fst link_rank] := snd link_rank;
    }
    ranks :=
      map((rank : Rat) => dampening * rank + (1 - dampening) / length(links),
          newRanks);
    iter := iter + 1;
  }
  return ranks;
}
\end{lstlisting}

\noindent
\rule{\linewidth}{.5pt}

\begin{lstlisting}[caption={\pagerank{} -- The final \mapreduce{} implementation},linebackgroundcolor=\colorinrange{12}{13}{codehighlight}, label={lst:pr-9}]
fn pageRank(links : [[Int]], dampening : Rat, iterations : Int) -> [Rat] {
  var iter : Int := 0;
  var ranks : [Rat] := replicate(length(links), 1 / length(links));
  while (iter < iterations) {
    var outRanks : [[Int] * Rat] := zip(links, ranks);
    var contribs : [Int * Rat] =
      flatMap((links_rank : [Int] * Rat) =>
                 map((link : Int) => (link,
                                      snd links_rank / length(fst links_rank)),
                     fst links_rank),
              outRanks);
    var rankUpdates : [Int * Rat] := reduceByKey((x : Rat) (y : Rat) => x + y, 0, contribs);
    var newRanks : [Rat] := replicate(length(links), 0);
    for (link_rank : rankUpdates) {
      newRanks[fst link_rank] := snd link_rank;
    }
    ranks :=
      map((rank : Rat) => dampening * rank + (1 - dampening) / length(links),
          newRanks);
    iter := iter + 1;
  }
  return ranks;
}
\end{lstlisting}

\section{Related Work}\label{sec:relatedwork}


A common approach to relational verification and program equivalence
is the use of product programs~\cite{productprograms}. Product
programs combine the states of two programs and interleave their
behavior in a single program. \emph{RVT}~\cite{rvt} proves the
equivalence of C programs by combining them in a product program. By
assuming that the program states are equal after each loop iteration,
\emph{RVT} avoids the need for user-specified or inferred loop
invariants and coupling predicates. 

Hawblitzel~et~al.~\cite{mutualsummaries} use a similar technique for
handling recursive function calls.
Felsing~et~al.~\cite{automatingregver} demonstrate that coupling
predicates for proving the equivalence of two programs can often be
inferred automatically. While the structure of imperative and
\mapreduce{} algorithms tends to be quite different, splitting the
translation into intermediate steps yields programs which are often
structurally similar. We have found that in this case, techniques such
as coupling predicates arise naturally and are useful for selected
parts of an equivalence proof.
De~Angelis et al.~\cite{DeAngelis16} present a further generalised
approach.

Radoi~et~al.~\cite{translatingimperative} describe an automatic
translation of imperative algorithms to \mapreduce{} algorithms based
on rewrite rules. While the rewrite rules are very similar to the ones
used in our approach, we complement rewrite rules by coupling
predicates.
Furthermore we are able to prove equivalence for algorithms for which
the automatic translation from Radoi~et~al. is not capable of
producing efficient \mapreduce{} algorithms.
The objective of verification imposes different constraints than the
automated translation -- in particular both programs are provided by
the user, so there is less flexibility needed in the formulation of
rewrite rules.

Chen~et~al.~\cite{sparkspecification} and
Radoi~et~al.~\cite{translatingimperative} describe languages and
sequential semantics for \mapreduce{} algorithms. Chen~et~al.\
describe an executable sequential specification in the Haskell
programming language focusing on capturing non-determinism
correctly. Radoi~et~al.\ use a language based on a lambda calculus as
the common representation for the previously described translation
from imperative to \mapreduce{} algorithms. While this language
closely resembles the language used in our approach, it lacks support
for representing some imperative constructs such as arbitrary
\emph{while}-loops.

Grossman~et~al.~\cite{equivalencespark} verify the equivalence of a
restricted subset of Spark programs by reducing the problem of
checking program equivalence to the validity of formulas in a
decidable fragment of first-order logic. While this approach is fully
automatic, it limits programs to Presburger arithmetic and requires
that they are synchronized in some way.

To the best of our knowledge, we are the first to propose a framework
for proving equivalence of \mapreduce{} and imperative programs.

\section{Conclusion}
\label{sec:conclusion}

In this paper we demonstrated how an imperative implementation of a
relevant, non-trivial algorithm can be iteratively transformed into an
equivalent efficient \mapreduce{} implementation.
The presentation bases on the formal framework described in~\cite{arxiv}.
Equivalence within this framework is guaranteed since the individual
applied transformations are either behaviour-preserving rewrite rules
or equivalence proofs using coupling predicates.
The example that has been used as a case study in this paper is the
\pagerank{} algorithm, a prototypical application case of the
\mapreduce{} programming model.  The transformation comprises eight
transformation steps.

Future work for proving equivalence between imperative and \mapreduce{}
implementations includes further automation of the transformation process.

\bibliographystyle{eptcs}
\bibliography{references}

\end{document}